\documentclass{elsart}

 \usepackage{graphics}
 \usepackage{epsfig}

\usepackage{amssymb}

\newcommand{\Da}{{\rm Da}}
\newcommand{\BE}{\begin{equation}}
\newcommand{\EE}{\end{equation}}
\newcommand{\BA}{\begin{eqnarray}}
\newcommand{\EA}{\end{eqnarray}}

\begin{document}

\begin{frontmatter}


\title{Filament bifurcations in a one-dimensional model of reacting excitable fluid flow}

\author[1]{Emilio Hern\'andez-Garc\'\i a\thanksref{email1}}
\thanks[email1]{\tt emilio@imedea.uib.es}
\author[1,2]{Crist\'obal L\'{o}pez\thanksref{email2}}
\thanks[email2]{\tt clopez@imedea.uib.es}
\author[3]{Zolt{\'a}n Neufeld\thanksref{email3}}
\thanks[email3]{\tt Z.Neufeld@damtp.cam.ac.uk}

\address[1]{Instituto
Mediterr\'aneo de Estudios Avanzados IMEDEA (CSIC-UIB)
\thanksref{www}, Campus de la Universitat de les Illes Balears,
E-07122 Palma de Mallorca, Spain.}
\thanks[www]{\tt http://www.imedea.uib.es/PhysDept/}
\address[2]{Departament de F{\'\i}sica, Universitat de les Illes
Balears, E-07122 Palma de Mallorca, Spain.}
\address[3]{Department of Applied Mathematics and Theoretical Physics,
University of Cambridge, Silver Street, Cambridge CB3 9EW, UK.}

\begin{abstract}
Recently, it has been shown that properties of excitable media
stirred by two-dimensional chaotic flows can be properly studied
in a one-dimensional framework \cite{excitablePRL,excitablePRE},
describing the transverse profile of the filament-like structures
observed in the system. Here, we perform a bifurcation analysis of
this one-dimensional approximation as a function of the {\it
Damk{\"o}hler} number, the ratio between the chemical and the
strain rates. Different branches of stable solutions are
calculated, and a Hopf bifurcation, leading to an oscillating
filament, identified.

\end{abstract}

\date{14 February 2003}
\end{frontmatter}


\section{Introduction}
\label{section:intro}

Chemically reacting substances in fluid flows are complex systems
important both from a fundamental point of view and for its
relevance in industrial and environmental contexts
\cite{issaos,Ottino94}. Typical incompressible chaotic flows
stretch fluid parcels along particular directions, with the
consequent contraction along the others. Thus, passive markers
draw filamental or lamellar structures that have been investigated
both from theory as from experiment\cite{Ottino88,Alvarez98}.
Chemical reactants are also stretched in this way, producing a
great increase in the surface of contact between different species
with deep effects on the global chemical kinetics\cite{Karolyi99}.
In addition to the strictly chemical interactions, biological
interactions in ecosystems (predation, grazing, consumption,
competition, ...) can also be described in the same framework, so
that stability and dynamics of aquatic ecosystems are also
strongly influenced by this filamentation
process\cite{PlanktonParadox}.

In this Paper we analyze a simplified model describing the
transverse chemical structure of these filaments, for the case in
which the chemical or biological activity is of the excitable
type. This includes in particular well known laboratory reactions,
such as the Belousov-Zhabotinsky reaction, but also reactions of
environmental importance, such as phytoplankton-zooplankton
competition\cite{Brindley}. Such kind of dynamics in
two-dimensional chaotic flows was analyzed in
\cite{excitablePRL,excitablePRE}. There it was found that some of
the dynamic regimes and transitions between them can be understood
in terms of a simple one-dimensional model, of the kind already
considered in \cite{Martin2000,McLeod2002,ZoltanFocus,Kiss2003}
that focusses in the transverse structure of the filaments. It was
mentioned in \cite{excitablePRE} that, in addition to the
bifurcations considered there in detail, there was a range of
parameters for which a complex coexistence of solutions and
bifurcations occurred, that may be of relevance to understand
qualitative behaviors of the two-dimensional chaotic flows. The
present paper focusses in that regime. We use the FitzHug-Nagumo
model as a prototypical excitable dynamics, but we expect that the
same qualitative results will be also obtained for other chemical
or biological excitable schemes. Similar behavior has been
observed in an excitable model of oceanic plankton population in
\cite{GRL}.

\section{The one-dimensional filament model}
\label{section:model}

Given a set of kinetic laws describing the time evolution of a
number of chemical or biological concentrations $\{C_i({\bf
x},t)\}_{i=1,...,N}$ in a homogeneous system,
\BE
{d C_i \over dt} = k F_i(C_1,...,C_N) \ \
\label{kl}
\EE
($k$ is a global reaction rate), a general partial differential
equation model describing their spatiotemporal evolution under the
simultaneous additional effects of advection in an incompressible
velocity field ${\bf v}({\bf x},t)$, and diffusion with diffusion
coefficients $\{D_i\}$, is the one given by the following
advection-reaction-diffusion equations:
\BE
{\partial C_i({\bf x},t) \over \partial t} + {\bf v}({\bf x},t)
\cdot \nabla C_i({\bf x},t) = k F_i(C_1,...,C_N) + D_i \nabla^2
C_i({\bf x},t)\ \ .
\label{ard}
\EE

In the following we restrict for simplicity to the case of equal
diffusion coefficients $D_i=D,\ \forall i$. In the immediate
vicinity of filaments or lamelae, one can approximate the flow by
its linearization around a point co-moving with a fluid element.
 In  areas  where the advection dynamics is hyperbolic this linearization
identifies principal directions $\{x_j\}$ along which the
corresponding velocity components read $v_j=\lambda_j x_j$.
Positive and negative values of the strain rate $\lambda_j$
identify expanding and contracting directions, respectively. Along
the expanding directions, diffusion and advection cooperate and
the chemical concentrations are fastly homogenized, whereas strong
gradients build-up in the contracting directions. This is the
origin of filamental (one expanding direction) or lamellar (two
expanding directions) structures. Gradients in Eq. (\ref{ard}) can
be safely neglected except along the contracting directions. We
consider here the case of filaments in two-dimensional flows, or
lamellae in three dimensions, so that there is only one
contracting direction, of strain rate $\lambda_i \equiv -\lambda$,
with $\lambda>0$. In this case, (\ref{ard}) reduces to an
effective one-dimensional model for the transverse profile of the
chemical distributions. By measuring time in units of
$\lambda^{-1}$, and space in units of $\sqrt{D/\lambda}$, we
arrive at:

\BE
{\partial C_i(x,t) \over \partial t} -x {\partial \over
\partial x} C_i(x,t) = \Da F_i(C_1,...,C_N) + {\partial^2 \over \partial x^2}
C_i(x,t)\ \ .
\label{FilamentModel}
\EE

We have introduced a Damk\"{o}hler number $\Da=k/\lambda$ as the
ratio between the chemical and the strain rate.

Several limitations are inherent to (\ref{FilamentModel}). It is a
strictly local model valid close to (moving) hyperbolic trajectories.
 The linearization assumption will only be
justified if the full filament width is smaller than any structure
in the velocity field. This implies in particular a very small
diffusion coefficient so that we are deeply in the Batchelor
regime. Additional qualitative changes of behavior, not contained
in (\ref{FilamentModel}), are expected when increasing the
diffusion coefficient in (\ref{ard}). Curvature effects or, more
importantly, any interaction between filaments, are completely
neglected. Finally, the use of a constant $\lambda$ neglects any
fluctuations of the stretching rate.

The FitzHugh-Nagumo (FN) model consists in a dynamics of the type
(\ref{kl}) for two interacting species, of concentrations $C_1$
and $C_2$, and reaction terms:

 \BA F_1 &=& C_1(a-C_1)(C_1-1)-C_2,
\label{FN1} \\
F_2 &=& \epsilon (C_1 -\gamma C_2).
\label{FN2}
\EA

The FN model behaves excitably when $\epsilon \ll 1$ so that
there is a separation between the fast evolution of the active
component or {\sl activator}, $C_1$, and the slow evolution of
$C_2$, the passive one or {\sl inhibitor}. We concentrate in the
parameter values $\epsilon=10^{-3}$, $\gamma=3.0$, and $a=0.25$,
for which robust excitable behavior is obtained.

\section{Filament solutions}
\label{section:filaments}

\begin{center}
\begin{figure}
\epsfig{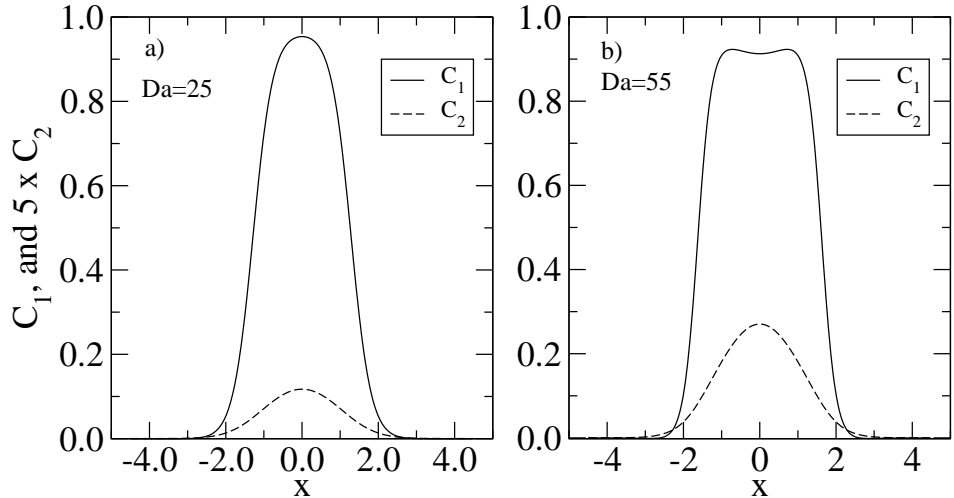} \caption{One-hump
solutions for the two chemical components: $C_1$ and $C_2$
(multiplied by a factor $5$); a) $\Da=25$; b)
$\Da=55$.}
\label{fig:onehump}
\end{figure}
\end{center}

The unexcited solution, $C_1=C_2=0$, is a linearly stable exact
solution of system (\ref{FilamentModel})-(\ref{FN2}). In addition,
the most notable solution is a pulse like steady solution,
centered on $x=0$, in which the activator is fully excited in a
central region, whereas the inhibitor remains at reduced
concentrations. This chemical distribution can be thought as the
transverse cut of one of the excited filaments seen in
two-dimensional excitable fluid flows. Pulse solutions of this type
are shown in Fig. \ref{fig:onehump}.

The width, $w_s$, of this solution is determined by the
competition between the tendency to expand, due to the joint
effects of the excitable chemistry and diffusion, and the
compression by the flow, and can be estimated
\cite{excitablePRL,excitablePRE} to be, for small $\epsilon$,
$w_s\approx (1-2a)\sqrt{\Da/2}$. It is very similar to what is
obtained in one-component bistable chemical
models\cite{ZoltanFocus}, so that the presence of the inhibitor
plays here only a secondary r{\^o}le.

As in bistable chemical models, this stable pulse solution
coexists in a range of parameters with a unstable pulse, of width
given at small $\epsilon$ approximately by \cite{excitablePRE}
$w_u \approx 2 /\sqrt{a\Da}$. When the value of $\Da$ decreases
(meaning that the chemistry becomes slower, or the stretching
faster) the two solutions approach, collide and disappear from
phase space in a saddle-node bifurcation \cite{excitablePRE}. This
happens at a value of $\Da_c \approx 2\sqrt{2/a} (1-2a)^{-1}$.
Thus, the excited pulse solution does not exists at small $\Da$,
and the existence of a critical $\Da_c$ explains
\cite{excitablePRE} the dynamical transition occurring in closed
two-dimensional chaotic flows between a situation in which local
perturbations of the unexcited state have limited impact on the
system, and a state in which they give rise to a global excitation
of the whole fluid. A related transition occurs in open flows
\cite{excitablePRE}.

\begin{center}
\begin{figure}
\epsfig{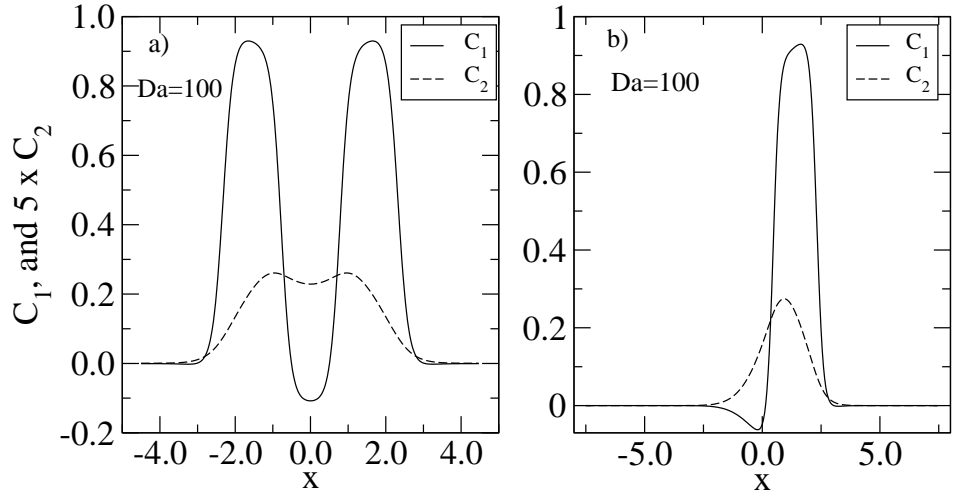} \caption{Two types of
pulses for $\Da=100$: a) The two-hump symmetric solution; b) the
asymmetric one.}
\label{fig:twohump}
\end{figure}
\end{center}

At larger values of $\Da$, the influence of the inhibitor becomes
more noticeable and, as a result, steady filament solutions have a
`two-humped' shape (Fig.\ref{fig:twohump}a). It was noted in
\cite{excitablePRE} that this kind of `double-filament', also seen
in two-dimensional flow simulations, can be regarded as a bound
state of two asymmetric counter-propagating excitable pulses,
which are also steady solutions of model
(\ref{FilamentModel})-(\ref{FN2}) coexisting with the symmetric
ones at large $\Da$ (Fig. \ref{fig:twohump}b). Of course, for each
value of $\Da$ there are two asymmetric solutions of this kind,
each one being the specular image of the other.

\section{Bifurcation behavior at intermediate Da}
\label{section:bifurcations}

\begin{center}
\begin{figure}
\epsfig{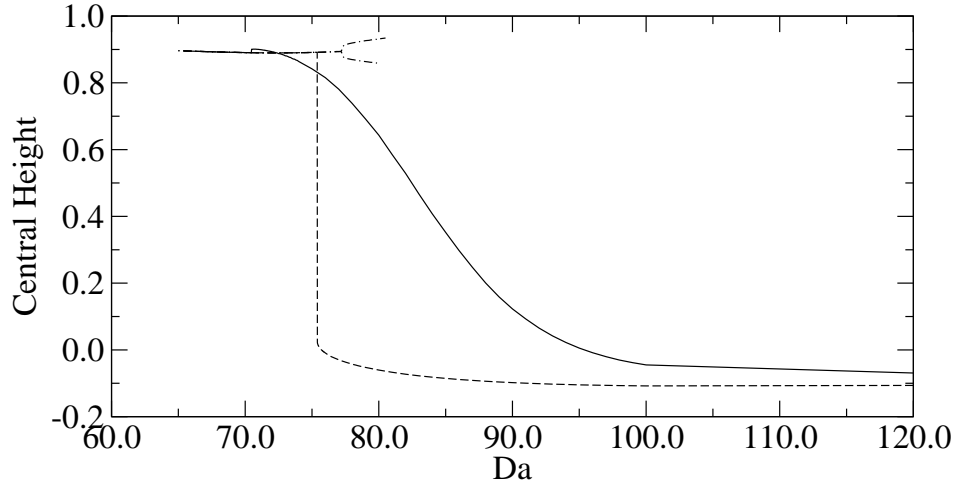} \caption{Bifurcation
diagram for the system given by Eq. (\ref{FilamentModel}). We plot
the height of stable solutions in terms of $\Da$. Solid line
corresponds to decreasing the value of $\Da$ from an initial state
of the asymmetric type, like that in Fig. \ref{fig:twohump}b), at
large $\Da$. Dashed line is obtained in the same way but beginning
from a symmetric two-hump pulse (Fig.\ref{fig:twohump}a).
Dashed-dotted line is obtained increasing the value of $\Da$ from
a symmetric one-hump pulse (Fig.\ref{fig:onehump}a)).
}\label{fig:bifurcaciones}
\end{figure}
\end{center}

It was noted in \cite{excitablePRE} that a complex bifurcation
scenario occurred between the small and the large values of $\Da$
for which the solutions in Figs. (\ref{fig:onehump}) and
(\ref{fig:twohump}) were respectively obtained. It happens that
these three solution branches are not directly connected, at least
for the values of $\epsilon$ and $\gamma$ considered here. We have
followed the three branches of steady stable pulse solutions by
starting with well developed solutions at large and small $\Da$,
and then performing slow changes in $\Da$. In this way we can only
follow stable solutions. A more complete characterization would
require also the continuation of unstable branches, that will be
performed elsewhere. We monitor the height of the activator at
$x=0$, and plot it in Fig. \ref{fig:bifurcaciones}.

\begin{center}
\begin{figure}
\epsfig{file=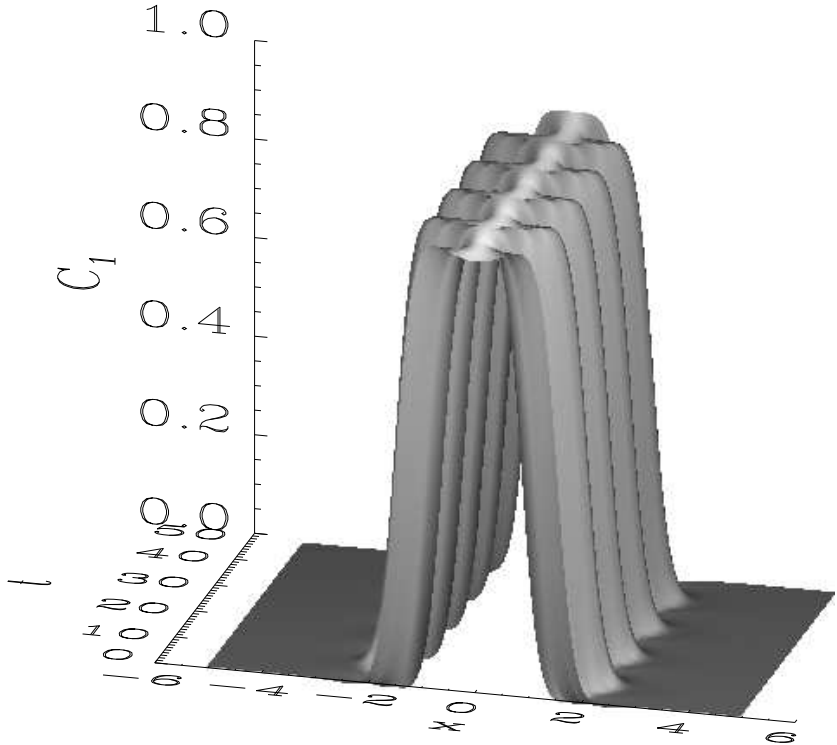,width=.9\linewidth} \caption{Spatio-temporal
plot of $C_1$ in an oscillating filament at
$\Da=79$.}\label{fig:oscillating}
\end{figure}
\end{center}

When increasing the value of $\Da$ on the symmetric one-humped
solution branch, the inhibitor value increases in the center, so
that for $\Da \gtrsim 45$, the central concentration of the
activator becomes a shallow minimum (Fig. \ref{fig:onehump}b). So,
above this parameter value, this branch can be called `slightly
two-humped', and it remains so until $Da \approx 77.1$. It is
however clearly different from the `strongly two-humped' solution
branch to which Fig. \ref{fig:twohump}a pertains. At $Da \approx
77.1$, a Hopf bifurcation occurs: the filament (both the $C_1$ and
the $C_2$ concentrations) pulsates in height, width and shape. We
show in Fig. (\ref{fig:oscillating}) this behavior. For $\Da$
above this Hopf bifurcation, we plot in Fig.
\ref{fig:bifurcaciones} the maximum and minimum central height
values attained during the oscillation. At $\Da \approx 81.0$
 the width becomes too narrow at some moment of the
oscillation and the filament collapses to the unexcited
($C_1=C_2=0$) solution, so that the limit cycle solution
disappears. This collapse probably reveals the collision of the
filament limit cycle with an unstable pulse solution that we have
not characterized. The basin of attraction of the oscillating
solution is rather narrow: at these values of $\Da$ it is easier
to be attracted by the two-hump symmetric steady solution branch
or the unexcited state.

On the other hand, if we start with a well developed two-hump
filament such as the one in Fig. \ref{fig:twohump}a at high $\Da$,
the two humps approach each other when decreasing $\Da$. Before
they fuse together, there is a discontinuous jump (at $\Da \approx
75.4$) to the steady `slightly two-humped filament' followed
before. This probably reveals a saddle-node bifurcation in which
the stable two-hump branch collides with an unstable one, that we
have not followed.

Finally, starting from the asymmetric filament of Fig.
\ref{fig:twohump}b, it approaches the $x=0$ axis when decreasing
$\Da$, becoming more and more symmetric. For the values of
$\epsilon$ and $\gamma$ used here, however, it does not join
smoothly with the symmetric branch in a forward pitchfork
bifurcation in which it would also join the other asymmetric
filament of opposite symmetry. Rather, it performs a (small)
discontinuous jump to the symmetric filament branch at $\Da
\approx 70.48$. This is probably the signature of a backward
pitchfork bifurcation, or of some other complex coexistence with
unstable solutions, that would also explain the observed range of
bistability between the symmetric and asymmetric filament
solutions.

\section{Summary}

We have investigated several pulse-like solutions of a
one-dimensional model intended to characterize the chemical
structure transverse to filaments or lamelae generated by chaotic
flows in reacting systems. The most striking feature has been the
finding of an oscillating solution, in which an activator-inhibitor
competition is sustained forever by the compressing strain. Moreover,
 parameter ranges have been found in which there is
coexistence of up to three stable pulse solutions, in addition to
the unexcited one. The transition behavior observed in
two-dimensional open and closed chaotic flows at intermediate
$\Da$ numbers \cite{excitablePRE} is probably related to this
complex phase space. Further work is needed to fully characterize
this relationship.

\section*{Acknowledgments}
We acknowledge support from MCyT (Spain) projects BFM2000-1108
(CONOCE) and REN2001-0802-C02-01/MAR (IMAGEN).

\end{document}